# Charge Retention in Lunar Regolith: Microscopic Processes and Constraints on Subsurface Charging

Ashley Daeun Jung[1,2], Hyun Jung Kim[2*]

[1]Department of Earth Science and Engineering, Imperial College London, London, United Kingdom

[2]STAR (Space Testing And payload sensoR) Lab, Department of Aerospace Engineering, KAIST, Daejeon, Republic of Korea

Corresponding author: Hyun Jung Kim hyunjung.kim1@kaist.ac.kr

## Key Points:

- Energy deposition, net charge transfer, and spatial charge separation constrain different aspects of lunar subsurface charging.
- Retained-charge responses link scales only under specific boundary, material, timing, and electrical-state conditions.
- Mean-field predictions and laboratory discharge evidence do not alone determine dielectric breakdown rates in lunar regolith.

## Abstract

Energetic-particle irradiation can produce persistent charge separation within lunar regolith, but the internal fields predicted by subsurface-charging models depend on how particle interactions are translated into retained charge. This critical review examines the evidence and assumptions connecting microscopic charge formation to macroscopic charging. We distinguish energy deposition, net charge transfer across a specified boundary, and the spatial distribution of signed excess charge. We organize these distinctions using an average retained-charge response defined for specified material, irradiation, electrical-state, and evaluation conditions. Its use as a volumetric source requires consistent charge accounting, compatible incident-flux definitions, and an adequate separation between response formation and subsequent transport. Applying these criteria to existing models, emission studies, and silicate irradiation experiments yields three assessments. First, emission measurements constrain electron loss across the measured boundary but do not locate internally redistributed charge. Second, evidence of internal reabsorption does not by itself establish that the charge separation sustaining subsurface fields is reduced; an illustrative calculation shows that the same net charge supply can produce different peak fields depending on transfer depths. Third, laboratory electrical damage in silicate crystals does not determine breakdown rates in cryogenic, porous regolith, which additionally require constraints on local fields, failure properties, and discharge recurrence. Joint constraints on boundary charge transfer, spatial charge separation, and post-irradiation relaxation are central to evaluating lunar subsurface-charging predictions.

## Plain Language Summary

The Moon has no thick atmosphere or global magnetic field, so high-energy particles from the Sun and deep space penetrate its soil, called regolith. In permanently shadowed craters near the poles, the soil is so cold that it conducts electricity extremely poorly, and electric charge left by these particles may remain for weeks. Earlier computer models predicted that large solar storms could build electric fields strong enough to make the soil break down electrically, like miniature lightning, gradually altering it. Other researchers questioned whether that much charge can build up in a loose, porous soil. We reviewed the existing evidence. Knowing how much energy the particles deposit, or how many electrons leave the soil surface, is not enough to know where positive and negative charge end up inside the soil. A simple calculation shows that electrons moving between different depths can change the predicted electric field even when the total amount of charge stays the same. Laboratory experiments show that irradiated mineral crystals can suffer electrical damage, but they do not yet tell us how often this happens in lunar soil. New measurements that track where charge sits inside the soil, and how quickly it disappears, are needed.

## 1 Introduction

The Moon lacks a substantial atmosphere and a global intrinsic magnetic field, leaving its surface exposed to solar radiation and charged particles from the surrounding space environment. Among these particles, solar energetic particles (SEPs) and galactic cosmic rays (GCRs) can penetrate beyond the immediate surface layer and interact with material within the lunar subsurface (Jordan et al., 2014, 2015). During SEP events, the Moon is exposed to intense fluxes of energetic electrons and ions (Halekas et al., 2009). GCRs originate from astrophysical sources and consist predominantly of protons and heavier nuclei, with a smaller electron component (Smart & Shea, 1985). The depths over which these particles interact with the regolith depend on their species and incident energies, together with the stopping properties and structure of the regolith (Berger et al., 2005; Geant4 Collaboration, 2024; Jordan et al., 2014).

The DC conductivity of lunar soil is extremely low and depends strongly on temperature (Olhoeft, Frisillo, & Strangway, 1974; Olhoeft, Frisillo, Strangway, & Sharpe, 1974). In permanently shadowed regions (PSRs), cryogenic temperatures are expected to further reduce conductivity, with existing models estimating dielectric relaxation times of several weeks (Jordan et al., 2014). Such prolonged relaxation allows charge to persist after the incident particle flux has declined, while continued irradiation can produce accumulation when charge supply exceeds redistribution or loss (Jordan et al., 2014). Spatially separated charge can generate internal electric fields, and previous modelling has suggested that large SEP events may produce fields capable of inducing dielectric breakdown and modifying the regolith (Jordan et al., 2014, 2015). Understanding the formation and persistence of subsurface charge is therefore important for interpreting the electrical behaviour and possible evolution of lunar regolith.

The lunar regolith is a heterogeneous, granular medium containing mineral grains, glassy components, and pore spaces (Haskin & Warren, 1991; McKay et al., 1991; Papike et al., 1982; Slabic et al., 2024). Within this structure, charge can be redistributed without necessarily leaving the material. For example, an electron transferred between neighbouring grains leaves the emitting grain more positive and the receiving grain more negative, while preserving the total charge of a volume containing both grains (Gandhi et al., 2027; Wang et al., 2016). The net charge of that volume alone therefore does not specify the spatial separation of positive and negative charge within it. Interpreting the electrical consequences of irradiation requires understanding how microscopic particle–material interactions produce these spatial charge distributions.

Existing macroscopic models provide a basis for investigating subsurface charging. Jordan et al. (2014) represent charging through idealized planar charge layers, whereas Sana and Mishra (2026) use a volumetric charging model. Both models predict strong subsurface fields, but whether such fields cause dielectric breakdown in lunar regolith remains debated. Hapke (2022) argued that secondary-electron transport through porous regolith could promote neutralization and escape before strong fields accumulate. MacLeod et al. (2023) demonstrated electrical discharges and microstructural damage in silicate samples under laboratory irradiation. These experiments support dielectric breakdown as a possible mechanism of silicate modification, but its occurrence in cryogenic, porous lunar regolith remains unresolved.

These studies address different parts of the connection between charge accumulation and material failure. Interpreting them together requires distinguishing evidence of electrical damage from evidence about the charge distributions that produce it. The long-term question motivating this work is what signed charge remains in lunar regolith after energetic-particle interactions, where it resides, and how it evolves over time.

Jordan (2022) reviewed how laboratory charging conditions affect the interpretation of charged-particle irradiation experiments and their implications for space weathering. The present review addresses a complementary step: how particle interactions are translated into the retained charge assumed by lunar subsurface-charging models. Its contribution is threefold. First, it distinguishes energy deposition, net charge transfer across a specified boundary, and the spatial distribution of retained signed charge, and defines the conditions under which an average retained-charge response can serve as a macroscopic charge source (Section 4). Second, it applies these conditions to assess what each class of evidence supports and what it leaves undetermined (Section 5, Table 2, and Figure 3). Third, it separates the inferential steps from predicted mean fields to local breakdown and cumulative weathering, and identifies the evidence required at each step (Section 6). Particular attention is given to the boundary, spatial scale, evaluation time, material conditions, and subsequent charge transport that determine whether laboratory and theoretical results can be applied to porous, low-conductivity lunar regolith.

This critical review examines lunar subsurface-charging models, irradiation experiments on lunar soils and relevant silicates, electron-emission and intergrain-charging studies, and research on charge transport and relaxation. The assessment focuses on the quantities measured or calculated, the boundaries and conditions they represent, and the inferences they support about retained charge and its electrical consequences. Studies of dayside surface charging and plasma sheaths are considered where they inform charge transfer or redistribution; studies of other materials are used only as explicitly identified mechanistic analogies. Statements about missing constraints refer to the literature reviewed here and do not imply that the corresponding physical processes are absent.

Section 2 compares the source, transport, and boundary assumptions of existing macroscopic models. Section 3 summarizes the microscopic processes needed for the subsequent assessment. Section 4 defines the retained-charge response and the conditions under which it can be used as a macroscopic source. Section 5 applies these conditions to the available evidence, and Section 6 examines the implications for internal fields, dielectric breakdown, and regolith weathering.

## 2 Macroscopic Descriptions of Lunar Subsurface Charging

Particle-induced charge distributions depend on both particle penetration and the subsequent transport, escape, and retention of charge carriers. Modelling the internal electric field therefore requires a description of where positive and negative charge resides and how this distribution evolves over time. The macroscopic models considered in this review describe this evolution through particle-induced charge supply and electrical relaxation within the regolith (Jordan et al., 2014; Sana & Mishra, 2026). Under a homogeneous, Ohmic approximation with constant material properties, the characteristic dielectric relaxation time is $\tau = \varepsilon/\sigma$, where $\varepsilon$ is the permittivity and $\sigma$ is the electrical conductivity. Charge can accumulate when energetic-particle irradiation supplies retained charge more rapidly than it can be redistributed or dissipated. Lower conductivity slows the relaxation of charge separation and the associated electric fields. In cryogenic PSRs, the low bulk conductivity assumed in existing modelling gives dielectric relaxation times of several weeks, allowing charge to persist after the incident flux has declined (Jordan et al., 2014). The competition between particle-induced charge supply and electrical relaxation provides a common basis for comparing these models, while leaving a central question: how is the charge supplied by particle irradiation determined?

To model charge accumulation within PSRs, Jordan et al. (2014) proposed a one-dimensional, time-dependent model using an idealized charge-layer representation. For SEP irradiation, the model represents proton and electron inputs through a shallower layer receiving positive charge and a deeper layer receiving negative charge. Their separation generates an internal electric field, allowing the subsurface to be treated as a leaky parallel-plate capacitor. Gauss's law relates the accumulated layer charges to the electric fields, while conduction redistributes charge over time (Jordan et al., 2014). The layers are approximated as laterally infinite because the irradiated region is much wider than the depth of particle penetration.

Using CRaTER and ACE/EPAM particle-flux measurements, Jordan et al. (2014) estimated persistent GCR-driven fields of up to approximately 700 V $m^{-1}$ and transient fields of at least $10^6$ V $m^{-1}$ during large SEP events. These results suggest that large SEP events may supply charge faster than conduction relaxes the resulting distribution, potentially producing fields sufficient for dielectric breakdown. The implications for breakdown and regolith weathering are discussed further in Section 6.

While the layer model captures electric-field generation through charge separation, it represents particle-induced charging using discrete layers. Particle penetration instead spans a range of depths, motivating a continuous description of charging within the irradiated region (Jordan et al., 2014; Sana & Mishra, 2026). Sana and Mishra employ a volumetric charging model in which incident proton and electron currents, and the resulting charge density, vary continuously with depth. Particle spectra, penetration ranges, and incidence angles are used to construct depth-dependent particle currents. The authors couple the charge continuity equation with Poisson's equation to calculate charge-density and electric-field profiles, including redistribution through electrical conduction (Sana & Mishra, 2026). This formulation accommodates overlapping depth ranges of positive and negative charging contributions and resolves the macroscopic field within the irradiated region.

The model predicts persistent GCR-driven fields of approximately 115 V $m^{-1}$ during solar maximum and 320 V $m^{-1}$ during solar minimum, whereas large SEP events generate transient but intense subsurface fields of order $10^7$ V $m^{-1}$ (Sana & Mishra, 2026). Sana and Mishra examine electric-field profiles as a function of depth and identify locations where the predicted field exceeds the adopted dielectric breakdown threshold. Their depth-dependent description therefore permits assessment of where field peaks develop within the charging region.

Differences in predicted field strengths cannot be attributed to spatial representation alone. Irradiation inputs and other model assumptions must also be considered. The principal distinction between these two models is how particle-induced charging is represented spatially and translated into an electric field. Jordan et al. (2014) track charge accumulated in effective layers and calculate the fields associated with their separation and combined net charge. Sana and Mishra (2026) instead calculate the evolution of a depth-dependent volume charge density from the spatial variation of particle and conduction currents. Both remain one-dimensional macroscopic descriptions, since resolving variation with depth does not itself resolve charge distributions within individual grains and pores. Table 1 compares their charge-source representations, material properties, boundary assumptions, and predicted quantities.

**Table 1**

*Comparison of Lunar Subsurface Charging Models*

| Comparison | Jordan et al. (2014) | Sana and Mishra (2026) |
|---|---|---|
| Spatial representation | 1-D; laterally infinite charge layers; principal formulation uses non-overlapping layers | 1-D; continuous depth-dependent volumetric charge density |
| Charge source | Protons and electrons deposited at representative depths; non-overlapping proton and electron layers | Depth-dependent particle currents converted into a volumetric charge source |
| Environmental inputs | CRaTER GCR and ACE/EPAM SEP data | Analytical spectra and ACE/EPAM SEP data |
| Predicted quantities | Layer charge densities and subsurface electric fields | Depth-dependent volume charge density and electric-field profiles |
| Penetration / incidence | PSTAR/ESTAR in $SiO_2$; $2\pi$ sr with representative penetration angle $\theta = 60°$ | PSTAR/ESTAR in $SiO_2$; angular integration over $2\pi$ sr |
| Charge–field formulation | Layer charge-balance equations coupled to Gauss's law; Ohmic dissipation | Depth-resolved continuity equation coupled to Gauss/Poisson formulation; Ohmic dissipation |
| Initial / boundary conditions | Initially uncharged layers and zero electric fields; no conductive current through the surface | Initially $\rho(z, 0) = 0$, $E(z, 0) = 0$; surface boundary $E(0, t) = 0$ |

| Comparison | Jordan et al. (2014) | Sana and Mishra (2026) |
|---|---|---|
| Regolith parameters | $\rho_{soil}$ = 1.5 g cm$^{-3}$ | $\rho_{soil}$ = 1.5 g cm$^{-3}$ |
| Electrical properties | $\varepsilon_r$ = 2 (Olhoeft & Strangway, 1975); temperature-dependent Ohmic $\sigma$ based on an Apollo-soil relation; representative PSR $\sigma$ ~ 10$^{-17}$ S m$^{-1}$ | $\varepsilon_r$ = 2 (Olhoeft & Strangway, 1975); temperature-dependent Ohmic $\sigma$ based on an Apollo-soil relation; calculations examine $\sigma$ = 10$^{-17}$ S m$^{-1}$, with 10$^{-16}$ S m$^{-1}$ also explored |
| Charge relaxation | $\tau = \varepsilon/\sigma$; typical PSR $\tau$ ~ 20 days | $\tau = \varepsilon/\sigma$; ~20 days for 10$^{-17}$ S m$^{-1}$; ~2 days for 10$^{-16}$ S m$^{-1}$ |
| Secondary electrons | Not calculated; near-surface secondary-electron escape discussed as reinforcing positive charging; emission from incident SEP electrons considered negligible | Secondary-electron generation neglected based on low emission yield |
| Field predictions | Persistent GCR fields up to ~700 V m$^{-1}$; transient SEP fields ≥10$^{6}$ V m$^{-1}$; event gap peaks 8.8–16 MV m$^{-1}$ for the three events examined | Persistent GCR fields of ~115–320 V m$^{-1}$; SEP: depth-localized fields of order 10$^{7}$ V m$^{-1}$ |
| Validation status | Particle-flux inputs are observation-based; subsurface charge and fields are model predictions | SEP event inputs are observation-based; GCR/parametric inputs are modelled; subsurface charge and fields are model predictions |

The evidence supporting these models informs their irradiation inputs and material response, but does not directly validate the resulting retained-charge distributions. Spacecraft particle-flux measurements and particle-range calculations inform the representation of irradiation and penetration; converting these inputs into deposited charge or depth-dependent charging currents introduces additional assumptions (Jordan et al., 2014; Sana & Mishra, 2026). For example, the continuous slowing down approximation (CSDA) represents stochastic energy losses through the average rate of kinetic energy loss within the target material. The resulting range approximates the average path length travelled as a particle slows to rest and neglects energy-loss straggling (Berger et al., 2005). It provides a characteristic scale for primary-particle transport, but does not directly determine the depth profile of energy deposition or retained charge.

Laboratory electrical-property measurements inform relaxation, although applying a conductivity–temperature relationship measured for an Apollo soil sample to PSR regolith requires assumptions about its material and spatial applicability (Jordan et al., 2014). Both models describe conduction using an Ohmic approximation (Jordan et al., 2014; Sana & Mishra, 2026). The particle-flux and electrical-property measurements therefore support different parts of the calculation: the former characterize the incident irradiation, while the latter inform the adopted relaxation behaviour. Neither directly measures the predicted subsurface charge or electric-field profile.

The models also differ in how they treat processes that can modify the charge supply before bulk conduction acts. Jordan et al. (2014) discuss near-surface secondary-electron escape as a possible reinforcement of positive charging without calculating secondary-electron transport, whereas Sana and Mishra (2026) neglect secondary-electron generation on the basis of a low emission yield. Hapke (2022) argued that secondary-electron emission and transport through porous regolith could inhibit the strong-field buildup predicted by Jordan et al. (2014). MacLeod et al. (2023) challenged the emission yield underlying this argument, stating that it was approximately an order of magnitude too high for SEP energies. These positions define the source assumptions under review: whether electron escape and internal reabsorption alter the depth-dependent charge supply prescribed in the models. The evidence bearing on these assumptions is assessed in Section 5.1.

The following section examines primary-particle outcomes and the transport, escape, reabsorption, trapping, and recombination of charge carriers. These processes provide the microscopic basis for assessing what signed charge remains within a specified regolith volume and how it changes over time.

## 3 Microscopic Processes Governing Retained Charge

The charge-source assumptions discussed in Section 2 require examination of what happens to incident particles and the charge carriers generated within the regolith. Their transport and subsequent interaction pathways determine how energy transfer gives rise to the charge remaining in the material. This section summarizes only the physics needed for the assessment in Section 5. Figure 1 summarizes these processes and their connection to the retained-charge response and macroscopic charging.

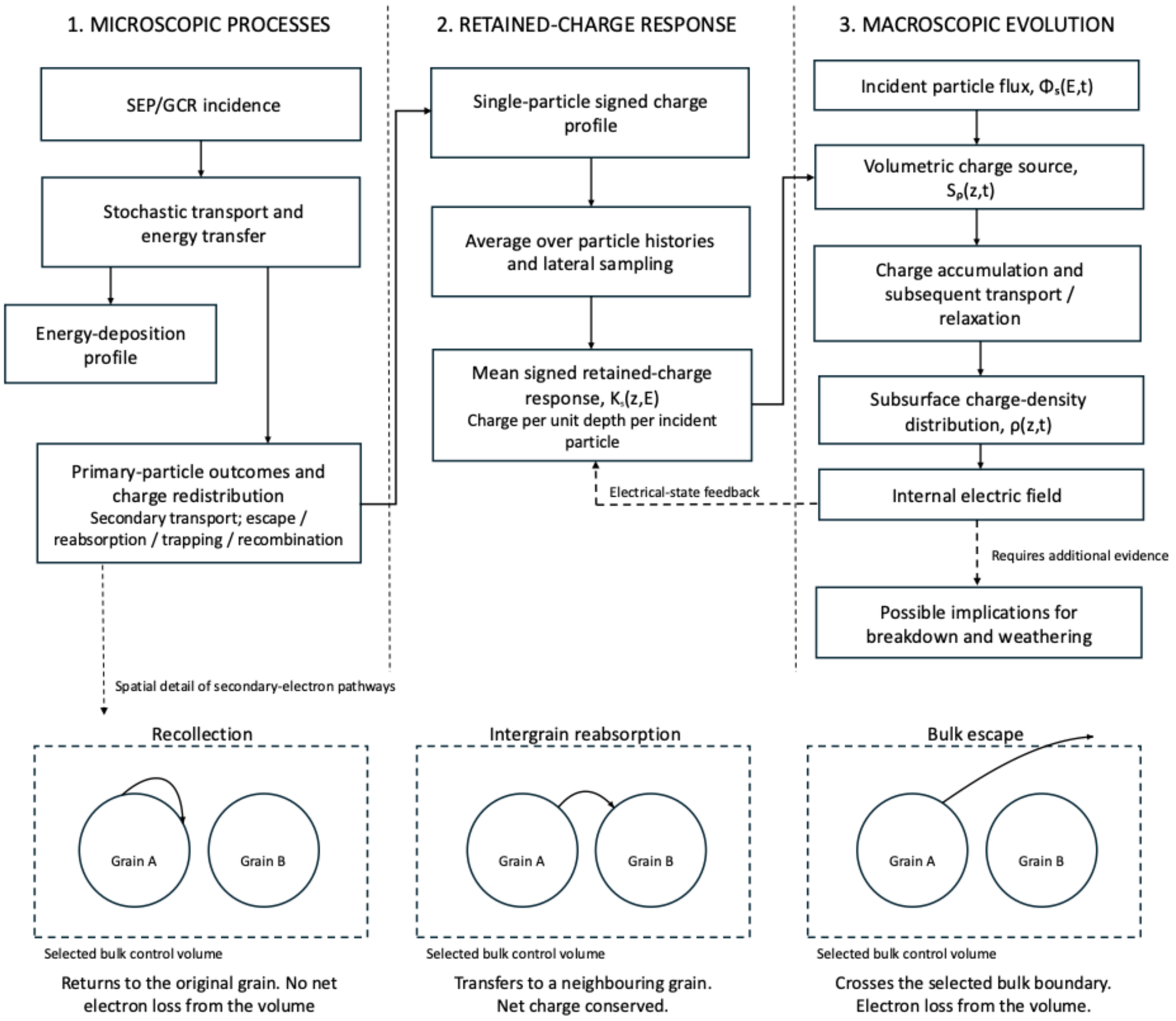

**Figure 1.** Conceptual framework linking microscopic energetic-particle interactions to macroscopic subsurface charging. The schematics represent recollection by the original grain, intergrain reabsorption, and escape across the selected bulk boundary.

### *3.1 Stochastic Transport, Energy Transfer, and Primary-Particle Outcomes*

When an energetic charged particle enters the regolith, its trajectory is governed by a stochastic sequence of scattering and energy-transfer interactions. The interaction probabilities depend on particle species and kinetic energy, along with the composition and atomic number density of the target material (Allison et al., 2016; Geant4 Collaboration, 2024; Nuclear Energy Agency, 2015). In a mineralogically heterogeneous regolith, the local interaction probability changes as the particle passes through different mineral grains and glassy components. Particles with identical initial energies and incidence angles may therefore follow different trajectories and experience different sequences of energy-transfer interactions (Agostinelli et al., 2003; Gandhi et al., 2027; Nuclear Energy Agency, 2015).

At the microscopic level, charged-particle interactions may be classified as elastic and inelastic (Berger et al., 2005; Nuclear Energy Agency, 2015). For incident electrons, elastic scattering primarily changes the direction of motion, with relatively little energy transferred to atomic recoil (Nuclear Energy Agency, 2015). Inelastic interactions instead transfer kinetic energy to the target through excitation and ionization. Excitation promotes a target electron to a higher-energy state, whereas ionization releases an electron from a bound state and leaves a positively ionized site or hole (Nuclear Energy Agency, 2015). The released electron may undergo further interactions within the material. At sufficiently high incident energies, radiative losses through bremsstrahlung also contribute to energy transfer (Berger et al., 2005; Geant4 Collaboration, 2024; Nuclear Energy Agency, 2015).

For protons, interactions with target electrons produce excitation and ionization, with the corresponding average energy loss per unit path length described by the electronic stopping power. Collisions with target nuclei transfer energy to recoiling atoms, potentially deflecting the proton and displacing target atoms; this contribution is described by the nuclear stopping power (Berger et al., 2005).

As these interactions accumulate, the primary particle may stop within a selected grain or regolith volume, return through the incident boundary by backscattering, or exit through another boundary by transmission (Gandhi et al., 2027; Nuclear Energy Agency, 2015). A backscattered or transmitted particle may deposit substantial energy before leaving, whereas a stopped particle may remain even when secondary electrons generated along its trajectory escape. In one particle-resolved simulation by Gandhi et al. (2027), a 1 keV solar-wind proton deposited approximately 950 eV in $SiO_2$ before backscattering and leaving the simulation volume. Most of its initial kinetic energy was transferred to the material, although the positive charge of the primary proton was not retained within the selected volume. This idealized keV-irradiation example illustrates the distinction between primary-particle energy deposition and primary-charge retention. However, it does not establish a quantitative response for lunar regolith under SEP or GCR irradiation.

Identifying whether the primary particle stops or leaves the selected volume is therefore one step in determining the retained charge. The full charge accounting also requires following the electrons and positive sites produced during energy transfer, including their subsequent redistribution, escape, trapping, and recombination.

### *3.2 Secondary Electron Transport and Charge Redistribution*

Secondary electrons originate from target electrons energized by incident-particle interactions (Seiler, 1982). Secondary-electron emission involves generation within the material, transport toward the solid–vacuum interface, and escape across the material surface barrier (Seiler, 1982). Generation alone does not guarantee emission: electrons may lose energy, recombine, or become trapped before reaching the interface (Dukes & Baragiola, 2013; Seiler, 1982). Only a fraction of the generated population reaches the surface with sufficient energy and an outward trajectory to cross the material surface barrier (Dukes & Baragiola, 2013; Seiler, 1982).

Crossing the solid–vacuum interface does not guarantee escape from the regolith. Local electrostatic potentials may return an emitted electron to its original grain, while neighbouring grains or pore walls may intercept and reabsorb it (Dukes & Baragiola, 2013; Wang et al., 2016). For an electron leaving a grain at a positive potential relative to its surroundings, escape requires sufficient kinetic energy to overcome the electrostatic potential barrier (Dukes & Baragiola, 2013). When an emitted electron is reabsorbed by a neighbouring grain, the emitting grain becomes more positive and the receiving grain more negative, producing intergrain charge separation without electron loss from a volume containing both grains (Gandhi et al., 2027; Wang et al., 2016). Wang et al. (2016) attributed the reduced secondary-electron emission yield of dusty surfaces to reabsorption by neighbouring particles or structures. The measured yield can also change as irradiation alters the sample's electrical state and the electrostatic conditions governing electron escape. To minimize these changes caused by accumulated charge, Vega et al. (2026) used low-fluence electron pulses and charge neutralization between pulses when measuring electron escape from lunar regolith simulants. Their yields therefore constrain external electron loss under conditions designed to minimize charging.

A measured secondary-electron yield constrains transfer across the measured boundary, but not the locations of reabsorption or the charge separation inside a larger regolith volume. The consequences of this distinction for the source assumptions of Section 2 are assessed in Section 5.1.

Recombination occurs when an electron fills a hole, removing an electron–hole pair and reducing charge separation (Nuclear Energy Agency, 2015). Electron–hole pair generation and recombination conserve net charge within a volume containing both carriers, although they can alter the spatial charge distribution. Trapping, in contrast, can restrict carrier transport and contribute to charge persistence. In the literature reviewed, trapping probabilities and lifetimes have not been constrained for lunar regolith under SEP or GCR irradiation. Studies of other amorphous materials (e.g., Konstantinou et al., 2022; Tran et al., 2023) indicate only that carrier localization is material dependent and cannot be inferred from structural disorder alone; they are not direct evidence for regolith. Microscopic trapping should therefore be treated as a plausible but insufficiently constrained contribution, distinct from the effective bulk relaxation described in Section 2.

### *3.3 Retained Charge*

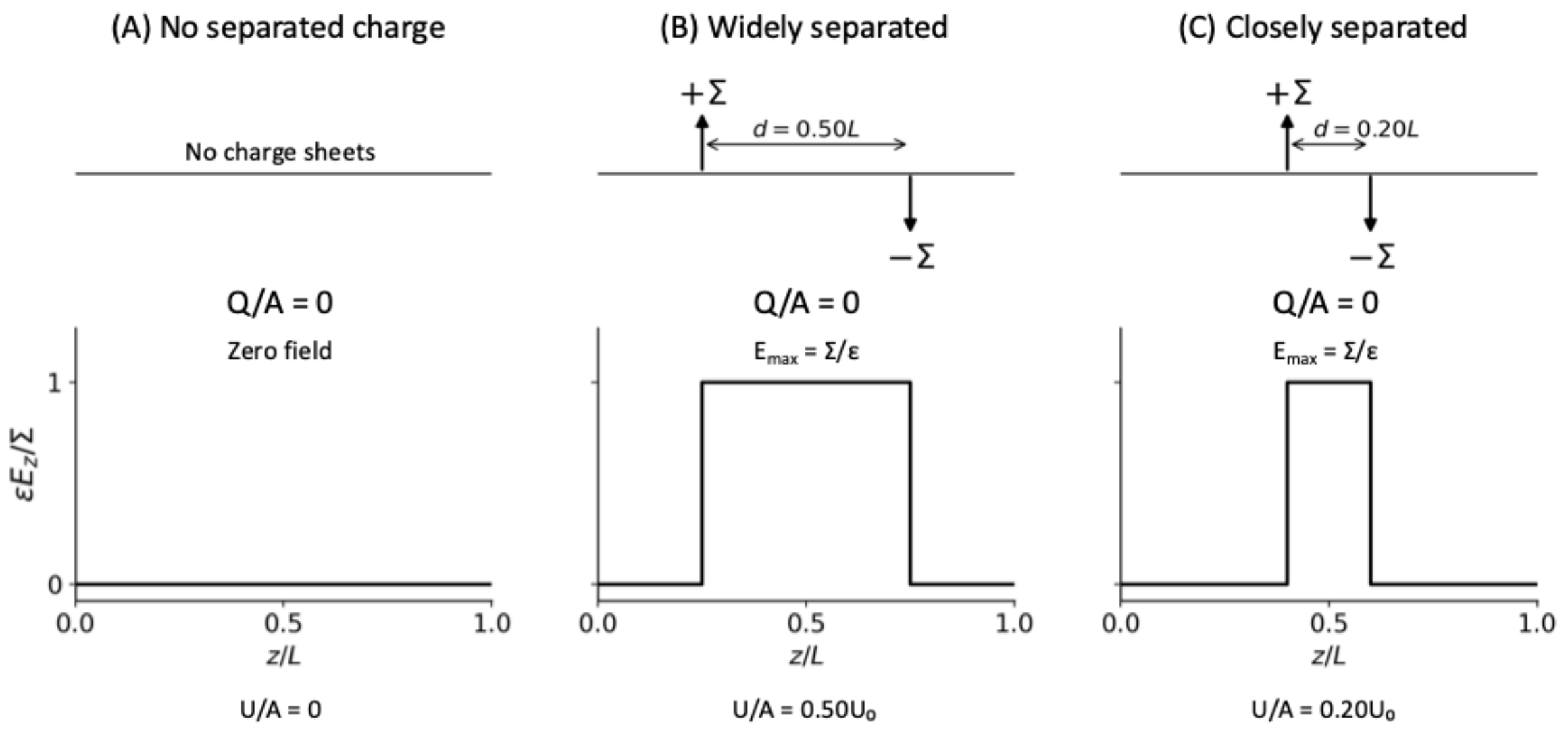

**Figure 2.** Idealized zero-net-charge configurations: (A) no charge sheets; (B, C) laterally infinite sheets with surface charge densities ±Σ, separated by 0.5*L* and 0.2*L*. Uniform permittivity *ε* and $E_z(0) = 0$ are assumed. B and C have identical peak fields Σ/*ε* but different field extents and stored energies. *L* is the domain depth, *A* is the lateral area, *U* is the stored electrostatic energy, and $U_0 = \Sigma^2 L/(2\varepsilon)$ is the reference energy per unit area.

The retention of an incident electron is commonly represented in grain-charging models as electron sticking. If an incident electron loses its kinetic energy and remains associated with the grain, it contributes negative charge, whereas secondary-electron emission removes negative charge from the grain and shifts its charge in the positive direction (Abbas et al., 2010). Sticking and emission are therefore competing grain-scale charging pathways.

However, neither process determines the charge retained within a bulk-regolith control volume. Following the processes described in Section 3.2, residual positive and negative charges may remain on different grains and at different locations within the regolith. Since these residual charges may be spatially separated, the outcome of one incident particle cannot always be described by a single net retained-charge value. The retained-charge distribution refers to the spatial distribution of excess positive and negative charge remaining within a specified control volume at a specified observation time, relative to the background charge distribution at that time. It may include contributions associated with a stopped or implanted primary particle, electron–hole separation, and charge separation produced by electron transfer between neighbouring grains. In other words, the retained-charge distribution describes where positive and negative excess charges remain, whereas net retained charge is their signed sum over the selected volume. Figure 2 illustrates how charge distributions with the same net charge can produce different internal electric-field profiles and stored electrostatic energies.

The selected boundary determines whether a particle's escape represents charge loss or internal redistribution. An electron leaving its grain of origin may remain within a larger regolith volume, while crossing that volume's outer boundary contributes to its net charge balance. Observation time must also be specified because continued transport, recollection, recombination, and relaxation can modify the charge distribution following irradiation. The retained-charge distribution therefore cannot be inferred directly from the energy-deposition profile (Gandhi et al., 2027; Geant4 Collaboration, 2024; Nuclear Energy Agency, 2015). Section 4 examines how these individual outcomes may be represented through an average retained-charge response and connected to macroscopic charging.

## 4 From Microscopic Charge Retention to Macroscopic Evolution

This section considers a conditional charge-source framework linking microscopic charge retention to macroscopic subsurface charging, as shown in Figure 1, specifying when retained charge is evaluated and under what conditions the resulting response can be used to describe continued irradiation. The definitions and conditions introduced here serve as the evaluation criteria applied to the literature in Section 5; they are not a new predictive model of lunar charging.

### *4.1 Defining the Average Retained-Charge Response*

For a single incident particle, the residual charges may remain on different grains and at different depths below the bulk-regolith surface (Gandhi et al., 2027). Since particles with identical initial conditions may undergo different stochastic interaction pathways (Agostinelli et al., 2003; Geant4 Collaboration, 2024; Nuclear Energy Agency, 2015), we consider the average signed charge remaining at each depth per incident particle as an input to macroscopic charging models.

In this framework, the average retained-charge response is the average signed excess charge per unit depth per incident particle of species $s$ and energy $E$. Depth $z$ is measured inward from the bulk-regolith surface. Excess charge is defined relative to the background at the same evaluation time, evolving from the same initial state without the additional incident particle:

$$\delta\rho_h(\mathbf{r}, t_h) = \rho_h(\mathbf{r}, t_h) - \rho_{\mathrm{ref}}(\mathbf{r}, t_h) \quad (1)$$

Here, $h$ denotes a particle history, $\mathbf{r}$ denotes position, and $t_h$ is the elapsed time after incidence at which the response is evaluated. The same evaluation time is used for the histories included in the average. $\rho_h$ is the charge density including the particle interaction, and $\rho_{\mathrm{ref}}$ is the background reference. The charge densities used here exclude polarization charge, which is already represented through the permittivity of the macroscopic model. If the background remains unchanged during the evaluation interval, the initial charge distribution provides the same reference.

For a selected regolith volume with lateral cross-sectional area $A$, the average retained-charge response is defined as:

$$K_s(z, E; t_h, \mathcal{X}) = \langle \iint_A \delta\rho_h(x, y, z, t_h)\, \mathrm{d}x\, \mathrm{d}y \rangle_h \quad (2)$$

The brackets denote averaging over particle histories under the specified conditions, including particles that stop, backscatter, or transmit. At each depth, the lateral integral sums the excess charge density over the selected area to obtain the excess charge per unit depth for one particle history. The area A defines the lateral extent of the selected control volume. The integral is not divided by A because the response represents charge per unit depth per incident particle. Integrating the averaged response over depth gives the mean excess net charge per incident particle (Equation 3), while combining it with the compatible incident particle flux gives a volumetric charge-source rate (Equation 4). For calculations using discrete depth intervals, called depth bins, this integral is represented by the excess charge in each bin divided by its thickness. These profiles are then averaged over particle histories. The resulting response has units of C $m^{-1}$ per incident particle. Positive and negative contributions are summed with their signs within each depth interval. This preserves charge separation between depths even when the net charge of the entire volume is zero.

$\mathcal{X}$ represents fixed conditions such as material properties, the incidence-angle distribution, and the initial electrical state. Particle species, energy, and regolith properties are the principal comparison variables. To isolate material effects, irradiation and initial electrical conditions should be matched, together with the evaluation time, control-volume boundaries, depth intervals, and sampling procedure. Simulants or idealized mineral configurations could provide such responses, although their applicability to lunar regolith would require separate assessment.

For a control volume extending from $z = 0$ to $z = L$ over the same lateral area $A$, integrating the response over depth gives the mean excess net charge per incident particle:

$$\int_0^L K_s(z, E; t_h, \mathcal{X})\, \mathrm{d}z = \langle \delta Q_{V,h}(t_h) \rangle_h \tag{3}$$

Here, $\delta Q_{V,h}$ is the volume-integrated excess charge for history $h$. With the same initial state and background reference, this change must equal the time-integrated net charge entering through all control-volume boundaries relative to the reference case. Charge accounting must therefore include incident and outgoing particles, together with any electrode or leakage currents present.

### *4.2 Selecting the Evaluation Time*

The evaluation time determines which processes are included in the retained-charge response. Primary-particle stopping does not mark the completion of charge formation, since secondary transport and subsequent redistribution can continue to modify the charge distribution (Seiler, 1982). The framework requires a time interval in which further changes from the initial interaction become sufficiently small, while bulk relaxation and irradiation-induced changes in the background remain small enough to treat the initial state as approximately fixed. Stability in most particle histories is insufficient if the remaining histories can still substantially change the average signed profile. Section 5.3 considers how the available evidence bears on this condition.

The dielectric relaxation time provides a bulk reference under the homogeneous, Ohmic approximation discussed in Section 2 (Jordan et al., 2014). An acceptable interval depends on the material, irradiation conditions, and accuracy required of the resulting charge and field predictions. If the initial retained-charge profile continues to change over timescales on which bulk relaxation or background evolution is already significant, its formation cannot be treated as a separate, completed input to subsequent macroscopic evolution. A time-dependent response or a coupled description of charge formation and macroscopic evolution would then be needed.

### *4.3 From the Average Response to a Volumetric Charge Source*

If this depth-dependent retained-charge response were known, it could be combined with the corresponding incident particle flux to estimate the volumetric rate at which charges are supplied as a function of depth. The formulation assumes a fixed retained-charge response and negligible changes in the incident flux, energy spectrum, and background electrical state during response formation, consistent with the timescale separation discussed in Section 4.2. Under these conditions, the volumetric charge-source rate is given by:

$$S_\rho(z, t) = \Sigma_s \int \Phi_s(E, t)\, K_s(z, E)\, \mathrm{d}E \tag{4}$$

where $\Phi_s(E, t)$ is the differential incident number flux entering through the reference surface per unit area, time, and energy, and $S_\rho(z, t)$ is the signed volumetric charge-source rate. With $K_s$ expressed in charge per unit depth per incident particle, the resulting source has units of C m$^{-3}$ s$^{-1}$. The flux and response must represent the same incidence distribution. Any angular projection already included in the surface-crossing flux must not be applied again in the source expression. Equivalently, $K_s$ must be normalized per particle crossing the reference surface, and its depth dependence must already reflect the angular distribution of those particles. The illustrative calculation in Section 5.5 adopts this convention (Equation 5).

The response $K_s$ must first be determined through measurements or particle-resolved modelling under the specified conditions. Equation 4 then weights these responses by the incident particle flux and sums their contributions over species and energy to obtain the charge-source rate. This rate describes charge supply, rather than the accumulated charge density. Subsequent evolution can be described through charge continuity and electrical transport, with Poisson's equation relating the charge distribution to the electric field, as in existing macroscopic descriptions (Jordan et al., 2014; Sana & Mishra, 2026).

Transport and redistribution caused by the particle interaction up to the evaluation time are included in $K_s$, while subsequent evolution is assigned to the macroscopic transport model. The same charge transfer must therefore not be counted in both descriptions. Charge conservation requires consistent treatment of charge crossing the selected control-volume boundary, as specified in Section 4.1. Whether the signed profile alone supplies sufficient information for subsequent transport is examined in Section 5.2.

### *4.4 Electrical-State Dependence and Reuse of the Response*

The retained-charge response may change as irradiation modifies the target's electrical state. Accumulated charge can alter local potentials and thereby affect escape, recollection, and redistribution. Self-consistent coupling between particle transport and evolving electric fields is already incorporated in the particle-resolved framework of Gandhi et al. (2027), demonstrated for representative lunar dayside grain-charging conditions. This provides a precedent for treating electrical feedback, although it does not establish the range over which a fixed response remains adequate for SEP- or GCR-driven subsurface charging.

The background may remain nearly unchanged during one particle interaction but change substantially as irradiation continues. Vega et al. (2026) observed a decrease in total electron yield during repeated 400 eV irradiation of an LHS-1 simulant sample without charge neutralization between pulses. The yield approached one emitted electron per incident electron, consistent with charging that increasingly limits electron escape. This result supports reassessing a fixed emission yield as charge accumulates, since a value measured with minimal charging may not represent electron loss at later electrical states.

If accumulated charge alters the response to later particles, a source constructed by repeatedly adding the same retained-charge response may become inaccurate. A fixed $K_s$ and additive source construction are therefore approximations within a defined range of material and electrical states. Changes in electrical state, such as local potential and trap occupation, should also be distinguished from irradiation-induced changes in material structure. The suitability of response reuse depends on how these changes affect the intended charge and field predictions. Section 5.3 considers how the available evidence bears on the range over which a fixed response remains appropriate.

## 5 Evidence Assessment and Limits of Macroscopic Predictions

The framework described in Section 4 identifies the microscopic response needed to construct a volumetric charge source. This section applies it as an evaluation standard. For each class of evidence we ask four questions: what was actually measured or calculated; what conclusion that quantity supports; which additional assumptions are required to apply it to lunar subsurface charging; and what it leaves undetermined. Four recurrent issues receive particular attention: the transfer of low-energy surface-charging results to SEP- and GCR-driven deep charging; the difference between single-crystal or compacted samples and porous regolith; the distinction between an observation and the model used to explain it; and the distinction between a process not identified in the reviewed literature and one that is physically excluded. Table 2 summarizes the quantities measured or calculated in the studies considered here, the conclusions they support, and the assumptions that remain when applying them to lunar subsurface charging and material damage. Each subsection first presents the reported results and our assessment, and then the remaining constraints; suggestions for further work are confined to the end of each subsection.

**Table 2**

*Evidence and Limitations in Linking Particle Irradiation to Lunar Subsurface Charging and Material Damage*

| Study and evidence type | Quantities measured or calculated | Assessment used in this review | Limits of application |
|---|---|---|---|
| Jordan et al. (2014) – macroscopic model | Inputs: measured particle fluxes<br>Calculated: layer charges and time-dependent internal fields<br>Domain: laterally infinite subsurface charge layers, with conduction between layers | *Mean field – model prediction:* demonstrates charge accumulation and field generation from prescribed particle currents competing with electrical relaxation | Effective deposition and bulk conductivity are assumed; internal charge and field distributions are not directly observed |
| Sana and Mishra (2026) – macroscopic model | Calculated: depth-dependent charge density and electric-field profiles<br>Domain: a one-dimensional subsurface continuum with a surface field boundary condition | *Mean field – model prediction:* resolves charge overlap and field peaks within a depth-dependent charging model | Greater depth resolution does not demonstrate greater predictive accuracy. Differences from the layer model depend on irradiation inputs and other assumptions |
| Dukes and Baragiola (2013) – lunar soil experiment | Measured: total electron yield and energy distribution of electrons emitted from Apollo soil under 40–2000 eV electron irradiation | *Measured external electron loss:* constrains total electron escape from the irradiated sample, including secondary and backscattered electrons | Internal electron trajectories, reabsorption locations, and the retained-charge profile are not |

| Study and evidence type | Quantities measured or calculated | Assessment used in this review | Limits of application |
|---|---|---|---|
| | Boundary: escape from the lunar-soil sample into an external collection region | | resolved. Incident energies are far below SEP energies |
| Vega et al. (2026) – electron-emission experiments and modelling | Measured: total and backscattered electron yields; emitted-electron spectrum; yield evolution during repeated irradiation<br>Boundary: electrons escaping the simulant sample into the external collector | *Measured emission and state dependence:* constrains electron loss across the sample boundary under tested conditions. Yield changes during irradiation without intervening neutralization show that electrical-state evolution can alter this source contribution | Internal reabsorption locations and depth-dependent retained charges are not resolved. Experimental conditions do not represent cryogenic conditions, and incident energies (≤5 keV) are far below SEP energies. Yield decay does not measure bulk charge relaxation |
| Wang et al. (2016) – comparative irradiation experiments | Measured: dust trajectories and potential profiles<br>Estimated: grain charges and forces<br>Domain: a laboratory dust surface; microcavity reabsorption is represented in the explanatory model | *Transport – model-based interpretation:* supports a reabsorption-based interpretation of dust charging and mobilization through comparisons of observed motion and potentials | Reabsorption fractions and grain-scale charge distributions were not directly measured. Experiments used Mars simulants; applicability to lunar regolith remains unresolved |
| Gandhi et al. (2027) – particle-resolved model | Calculated: particle trajectories and outcomes, deposited-charge distributions, potentials, and electric fields<br>Domain: specified simulation geometry; distinguishes material-interface crossings from escape out of the simulation volume | *Source and transport – particle-resolved calculation:* demonstrates how particle retention, intergrain electron transfer, and escape can be linked to charge accumulation and electrical feedback | Deposited-charge mobility is neglected; leakage treated macroscopically; demonstrated for dayside cases; further validation for SEP/GCR conditions is required. |
| Hapke (2022) – theoretical critique | Estimated: SEP stopping depths, charges per grain, and electric-field magnitudes<br>Calculated: grain number density, mean grain size, intergrain spacing, and collision mean free path<br>Scope: grain-surface emission, electron transport through pore spaces, and escape from porous regolith | *Source and transport – theoretical critique:* identifies secondary-electron neutralization and escape in porous regolith as mechanisms that could limit charge buildup | Depends on the applicability of emission yields and rapid-neutralization assumptions. Does not calculate the evolving retained-charge profile |
| MacLeod et al. (2023) – electron-irradiation experiments | Measured/observed: discharge-associated current changes and surface/subsurface damage in individual crystals<br>Calculated: normal-incidence electron flux and fluence<br>Domain: irradiated individual silicate single crystals and their surfaces | *Local failure and damage – experimental evidence:* demonstrates that electron irradiation can produce surface flashovers, subsurface breakdown, and silicate damage under laboratory conditions | Individual silicate crystals do not represent a porous regolith assemblage. Irradiation used 30 keV electrons at room temperature. Experiments do not determine how intergrain electron transfer and pore geometry affect charge retention and discharge under lunar regolith conditions |

| Study and evidence type | Quantities measured or calculated | Assessment used in this review | Limits of application |
|---|---|---|---|
| Huang et al. (2021) – molecular dynamics simulation | Calculated: bond breaking and chemical products, including water formation during modelled breakdown<br>Domain: a nanoscale, hydrogen-bearing silica system subjected to an electric field | *Damage – molecular-scale simulation:* supports a possible chemical-modification pathway under prescribed field and material conditions | Simulations do not determine whether lunar charging can produce the conditions required for breakdown |

### *5.1 Constraints on Charge Supply and the Retained-Charge Response*

*Reported results.* Emission and dust-charging studies provide the most direct experimental constraints on the source assumptions identified in Section 2. Dukes and Baragiola (2013) measured total electron yields and emitted-electron energy distributions from Apollo soil under 40–2000 eV electron irradiation and interpreted reduced yields in terms of reabsorption within microscopic surface cavities. Vega et al. (2026) measured secondary and backscattered yields of lunar simulants and related granular insulators using low-fluence pulsed irradiation with charge neutralization between pulses, and showed that yields evolve when charge is allowed to accumulate. Wang et al. (2016) explained laboratory dust mobilization with a patched-charge model involving reabsorption of emitted electrons within intergrain microcavities, and Gandhi et al. (2027) obtained charge accumulation in such microcavities in particle-resolved simulations of dayside grain charging. MacLeod et al. (2023) argued that the secondary-electron yield adopted by Hapke (2022) was approximately an order of magnitude too high for SEP energies.

*Assessment.* These studies constrain different quantities. The emission measurements constrain electrons that cross the sample boundary into the collection region; they do not locate electrons reabsorbed within the material. In Wang et al. (2016), localized charges and electrostatic forces belong to the explanatory model rather than to a measured charge distribution, and the reabsorption in Gandhi et al. (2027) is a calculated outcome for dayside conditions. Together, these studies support including internal reabsorption among the processes that can affect retained charge, and they show that a low emitted yield is insufficient grounds for neglecting internal electron transfer. They do not establish that reabsorption reduces the charge separation responsible for a macroscopic subsurface field. That outcome depends on where electrons depart and arrive relative to the existing charge distribution, on the amount transferred, and on whether the transfer is fast enough to compete with continued supply; redistribution may reduce the separation between oppositely charged regions or create new localized separation (Section 5.5). The disagreement between Hapke (2022) and MacLeod et al. (2023) further shows that adopted emission yields must match the incident particle energies. Consequently, neither the measured emission yield nor evidence of internal reabsorption determines whether the depth-dependent charge separation prescribed in existing models is maintained.

*Scope limits.* Three limits apply when transferring this evidence to deep charging. First, the emission measurements used incident electrons of at most a few keV (Dukes & Baragiola, 2013; Vega et al., 2026) under non-cryogenic conditions, whereas SEP electrons and protons deposit charge at depths from micrometres to centimetres in cold PSR regolith; surface yields therefore constrain near-surface escape but not the deep source profile. Second, the trajectories that return emitted electrons to the regolith depend on grain and pore geometry and on local potentials, which differ among dust layers, compacted simulants, and undisturbed regolith. Third, in the literature reviewed we found no measurement of the depth-dependent retained charge in lunar soils or simulants under SEP- or GCR-relevant irradiation. This absence means that the depth profile is unconstrained, not that internal redistribution is either negligible or dominant.

*Remaining constraints on the full response.* The studies reviewed do not provide a complete quantitative description of how primary-particle outcomes and subsequent carrier processes contribute to the retained-charge response under lunar conditions. The dependence of these processes on mineralogy, grain and pore geometry, impact-glass content, and pre-existing electrostatic potential remains uncertain. Microscopic trapping illustrates the limits of transferring evidence between materials: because carrier localization is material dependent and cannot be inferred from structural disorder alone (Konstantinou et al., 2022; Tran et al., 2023), studies of other amorphous materials indicate possible trapping behaviour without constraining it in regolith. Responses from simulants or idealized mineral configurations would likewise constrain the tested materials; their use for lunar conditions would require assessment of the relevant material differences.

Which uncertain contributions most strongly affect the sign, magnitude, and depth distribution of retained charge therefore remains undetermined. Sensitivity calculations that vary these contributions within physically justified ranges could identify the measurements most needed, addressing the consequences of uncertainty rather than the underlying process probabilities. Irradiation of characterized simulants and lunar samples at cryogenic temperature and high vacuum (Jordan et al., 2014), under the matched conditions of Section 4.1, could then test which material differences change the response significantly.

### *5.2 Representativeness and Sufficiency of the Average Profile*

Under a laterally uniform electrostatic approximation with specified permittivity and boundary conditions, cumulative charge with depth constrains the mean vertical field. Local fields near grain contacts and pore walls require additional spatial information. A laterally averaged, depth-dependent description can therefore be adequate for one observable while remaining insufficient for another. Its adequacy must be assessed for a specified field quantity and spatial scale.

The macroscopic descriptions of Jordan et al. (2014) and Sana and Mishra (2026) connect charge distributions to fields within such one-dimensional representations. For the present framework, the profile retains depth dependence during lateral averaging. The macroscopic input does not have to preserve every grain-scale charge arrangement; however, it must capture the effects of microscopic processes on the mean signed profile.

In the evidence reviewed, the applicability of this averaged description to heterogeneous regolith under SEP or GCR irradiation has not been tested. Neither macroscopic model derives its source from a resolved retained-charge response, and the particle-resolved calculations of Gandhi et al. (2027) have not been reduced to a laterally averaged, depth-dependent profile for these irradiation conditions. The remaining question is whether the adopted averaging scale, effective permittivity, and transport description adequately represent the material for the mean-field prediction.

Reabsorption is an example of a microscopic contribution that could be represented through average quantities. As discussed in Section 5.1, Wang et al. (2016) support considering reabsorption in grain-scale charging but do not provide a quantitative retained-charge response for SEP or GCR irradiation. An average reabsorption fraction constrained under relevant conditions, combined with information about emission and reabsorption depths, could help represent this contribution without tracking every electron. The depth information matters, since transfer within one depth interval and transfer between intervals can produce different signed profiles.

Predicting subsequent evolution requires more than the charge profile at a given time. Jordan et al. (2014) and Sana and Mishra (2026) describe redistribution through bulk electrical conduction, assuming that the adopted conductivity represents transport at the modelled scale. An additional distinction between mobile and trapped carriers may be needed if their populations and exchange significantly affect charge evolution. A trapped population could be treated as approximately fixed only over an interval in which detrapping and loss are negligible.

Two tests would address these points. First, repeated measurements or particle-resolved simulations under the conditions of Section 4.1 could estimate the mean response and its variability, with statistical stability assessed by increasing the number of histories and the extent of lateral sampling; stability under additional sampling, however, does not show that the profile preserves the information needed for field prediction. Second, spatially resolved calculations could be compared with calculations using the corresponding depth-dependent average, with irradiation, boundary conditions, transport assumptions, permittivity, field component, averaging procedure, and spatial scale matched. Agreement in mean-field magnitude and peak depth within a stated tolerance would support the averaging approximation for those quantities within the compared models; independent evaluation of accuracy is addressed in Section 5.4. Comparisons that begin from matched signed profiles but differ in carrier state could similarly test whether an effective bulk transport model is sufficient.

### *5.3 Validity of the Evaluation Interval and Response Reuse*

A long bulk dielectric relaxation time does not necessarily mean that the initial retained-charge response has stabilized. The reviewed studies do not report the time required for a retained-charge profile to stabilize after SEP- or GCR-like irradiation, so the timescale separation assumed in Section 4.2 is unverified. Evidence on response reuse is limited but informative: the yield evolution reported by Vega et al. (2026) shows that the electron-loss contribution changes with electrical state under the tested conditions, whereas Gandhi et al. (2027) demonstrate self-consistent feedback only for dayside grain charging. A fixed response is therefore an unverified approximation for deep charging rather than an established property.

Following the same particle histories to later observation times, and comparing their average signed profiles with the evolving background charge and field, would test the evaluation interval. Little change over a finite interval cannot exclude later redistribution, so an interval set for one material and irradiation condition would require reassessment before application to others. Responses evaluated at different initial potentials or trap occupations could identify the range over which a fixed response remains adequate. Accumulated fluence or dose can track exposure but does not show whether the response has changed enough to matter; that significance should be judged by comparing predicted charge profiles, field magnitudes, and peak locations obtained with fixed and state-dependent responses, against a tolerance reflecting the intended prediction and input uncertainties.

### *5.4 Validation Through Charge and Field Predictions*

In the studies reviewed here, lunar subsurface-charging predictions have not been compared with independent measurements of subsurface charge or electric fields. The particle-flux and electrical-property data used by Jordan et al. (2014) and Sana and Mishra (2026) characterize model inputs; in the sense of Trucano et al. (2006), the predicted fields rest on observation-based inputs but have not been validated against measured outputs.

Different measurements test different parts of the framework. Under the material and irradiation conditions discussed in Section 5.1, comparing net charge transfer across the selected boundary with the change in net retained charge would test charge accounting. Even if these quantities agree, they do not specify where positive and negative charge resides within the volume. Spatially resolved charge or field measurements would address the predicted distribution, while time-resolved measurements of charge profiles and electric fields during and after irradiation would evaluate its evolution.

A useful model comparison would hold the incident irradiation, boundary conditions, and subsequent transport assumptions fixed while changing the charge-source description.

Comparing an effective deposition prescription with a source constructed from a constrained retained-charge response would show how the microscopic input affects the charge profile and the magnitude and location of field peaks. This comparison would isolate the effect of the source description within the adopted transport model, but would not test the transport assumptions themselves. A more detailed source description would therefore require comparison with independent measurements to demonstrate improved accuracy.

Measurements used to estimate source or transport parameters provide calibration rather than independent validation (Trucano et al., 2006). Predictions should therefore be compared with separate measurements that were not used to determine those parameters. Agreement with a field measurement alone would not prove whether the source and transport descriptions are individually adequate. Joint measurements of charge profiles, electric fields, and charge transfer across the selected boundary could help distinguish between alternative descriptions of charge supply and subsequent transport, with uncertainties in the source and electrical properties evaluated separately. Figure 3 summarizes the physical constraints provided by different measurement types, the scope of the interpretations they support, and whether each type of evidence has been reported under conditions relevant to the Moon. Its implications for dielectric breakdown and regolith weathering are considered in Section 6.

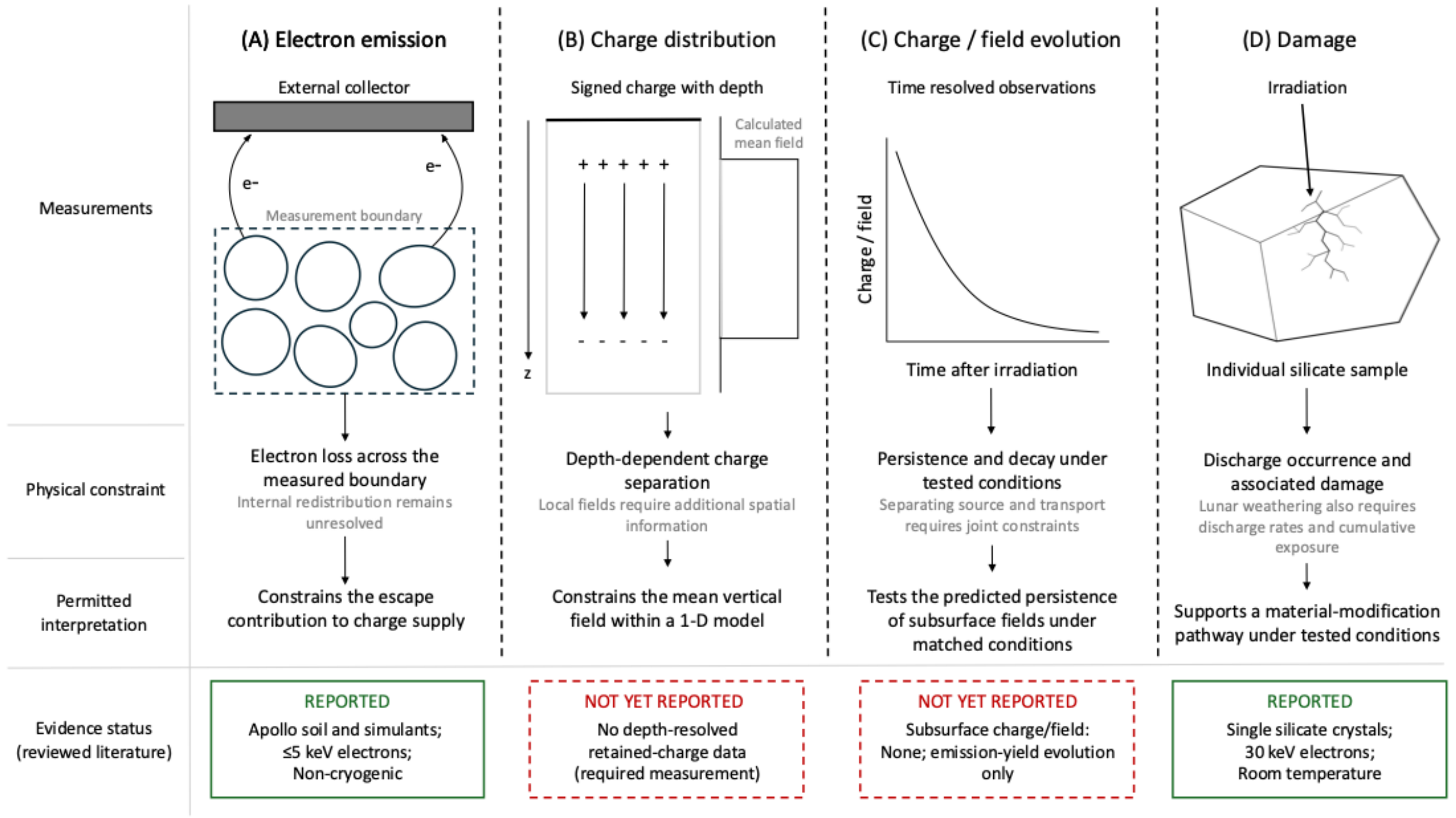

**Figure 3.** Measurement types, physical constraints, and permitted interpretations for (A) electron emission, (B) charge distribution, (C) charge and field evolution, and (D) material damage. Grey text identifies remaining limitations. Drawings and graphs are schematic. The bottom row gives the evidence status of each pathway in the literature reviewed here (solid boxes: reported; dashed boxes: not yet reported). (A) Reported for Apollo soil and lunar simulants with incident electrons of ≤5 keV under non-cryogenic conditions (Dukes & Baragiola, 2013; Vega et al., 2026). (B) Not reported for lunar soils or simulants; a required measurement. (C) Not reported for subsurface charge or fields; only the evolution of emission yield during repeated irradiation has been reported (Vega et al., 2026). (D) Reported for individual silicate crystals irradiated with 30 keV electrons at room temperature (MacLeod et al., 2023); not reported for porous regolith or cryogenic conditions.

### *5.5 Illustrative Sensitivity to Internal Charge Redistribution*

An illustrative calculation examines whether internal redistribution can alter a predicted mean field while preserving the net charge supply. We use a calculation based on the Gaussian SEP formulation of Sana and Mishra (2026), holding the incident irradiation, electrical properties, initial state, and boundary conditions fixed while modifying the charge source. This modification represents an assumed contribution to the retained-charge response. Its magnitude, departure, and arrival depths are prescribed for this specific sensitivity test and are not inferred from measurements or particle-resolved modelling.

The baseline uses equal incident proton and electron fluxes with peak values of $10^6$ $cm^{-2}$ $s^{-1}$, a Gaussian pulse centred on day 10 with a standard deviation of 2.5/3 ≈ 0.83 days, and particle energies of 0.01–10 MeV. The calculation adopts uniform permittivity $\varepsilon = 2\varepsilon_0$, conductivity $\sigma = 10^{-17}$ S $m^{-1}$, zero initial charge, and $E(0, t) = 0$. NIST $SiO_2$ range data (Berger et al., 2005) are converted to lengths using a bulk density of 1.5 g $cm^{-3}$. Electron CSDA ranges are used throughout, while proton CSDA and projected ranges are evaluated separately. With depth $z$ positive inward, the prescribed depth dependence of each particle current follows the range-based angular integration of Sana and Mishra (2026):

$$P_s(z) = \int_{E_{\min}}^{E_{\max}} w_s(E) \max[1 - z/R_s(E), 0]\, dE \quad (5)$$

$$J_{\text{particle}}(z, t) = e[\Phi_p(t)P_p(z) - \Phi_e(t)P_e(z)] \quad (6)$$

$$S_\rho^{(0)}(z, t) = -\partial J_{\text{particle}}(z, t)/\partial z \quad (7)$$

Here, $P_s(z)$ is the dimensionless depth profile of the particle current for species $s$, normalized to unity at the surface. $J_{\text{particle}}(z, t)$ is the signed electrical current density carried by the prescribed proton and electron populations, excluding the subsequent conduction current. The terms $e\Phi_p(t)P_p(z)$ and $-e\Phi_e(t)P_e(z)$ are the proton and electron contributions. The negative depth derivative of $J_{\text{particle}}(z, t)$ defines the baseline volumetric charge-source rate $S_\rho^{(0)}(z, t)$, where the superscript (0) denotes the baseline before the redistribution correction is added. As the particles penetrate into the material, a decrease in the inward proton current produces a positive source, whereas a decrease in the magnitude of the inward electron current produces a negative source.

The function $w_s(E)$ is the unit-normalized energy spectrum, $R_s(E)$ is the selected NIST range converted to a length, and $\Phi_s(t)$ is the energy-integrated incident number flux used to normalize the current. The constant $e$ is the positive elementary-charge magnitude. Following the spectral parameters adopted by Sana and Mishra (2026), the proton spectrum is specified as:

$$w_p(E) = C_p\, (E/E_0)^{-1.3} \exp[-E/(31\ \text{MeV})] \quad (8)$$

Here, $E_0$ = 1 MeV. The electron spectrum is specified as:

$$w_e(E) = C_e\ (E/E_\beta)^{-1.8},\ \text{for } E < E_\beta \quad (9a)$$

$$w_e(E) = C_e\ (E/E_\beta)^{-3.7},\ \text{for } E \geq E_\beta \quad (9b)$$

Here, $E_\beta$ = 0.06 MeV. The normalization constants $C_p$ and $C_e$ are chosen separately so that each spectrum integrates to unity over 0.01-10 MeV. Both electron branches equal $C_e$ at $E_\beta$, so the piecewise spectrum is continuous at the break energy.

The depth factor in Equation 5 follows from the assumptions of Sana and Mishra (2026). A particle of range $R_s(E)$ incident at angle $\vartheta$ from the inward surface normal is assumed to travel in a straight line and to reach depth $R_s(E)\cos\vartheta$, and incident directions are weighted by $\sin\vartheta\, d\vartheta$ over $0 \leq \vartheta \leq \pi/2$, that is, uniformly in $\mu = \cos\vartheta$. The fraction of particles of energy $E$ that cross depth $z$ is then the fraction with $\mu \geq z/R_s(E)$, which equals $\max[1 - z/R_s(E), 0]$. Because this weighting counts particles crossing the surface, rather than projecting an isotropic radiance onto the surface normal, $\Phi_s(t)$ is the surface-crossing number flux defined in Section 4.3 and no further $\cos\vartheta$ factor is applied. A cosine-weighted (isotropic-radiance) distribution would instead give $\max[1 - (z/R_s)^2, 0]$; we retain the uniform-$\mu$ weighting for comparability with Sana and Mishra (2026). The mapping neglects scattering and range straggling and serves only to construct the prescribed baseline for the sensitivity test.

The baseline source is coupled to charge continuity with Ohmic conduction and to Gauss's law,

$$\partial\rho(z, t)/\partial t = S_\rho(z, t) - \partial[\sigma E(z, t)]/\partial z \quad (10)$$

$$E(z, t) = (1/\varepsilon) \int_0^z \rho(z', t)\, dz' \quad (11)$$

Equation 11 incorporates the surface condition $E(0, t) = 0$. NIST range values were interpolated linearly in log–log space. Spectrum normalization and the energy integral in Equation 5 were evaluated using the composite trapezoidal rule on 12,000 logarithmically spaced energy points. With uniform electrical properties and a common Gaussian time dependence for the source, the field solution separates into spatial and temporal factors. The temporal factor was integrated using an adaptive Runge–Kutta method (RK45). The baseline spatial factor was evaluated directly from $P_e(z) - P_p(z)$, while the redistribution contribution introduced below was evaluated using cumulative lognormal distributions. Fields were evaluated at 6,000 logarithmically spaced depths from $10^{-9}$ to 0.05 m. The surface condition was incorporated at z = 0 through the spatial solution, rather than imposed at the first sampled depth.

Numerical convergence was assessed for the baseline and Cases B and C at f = 0.10, using both proton-range prescriptions. Increasing the depth and energy grids to 12,000 and 24,000 points, respectively, produced a relative change of less than $10^{-6}$ in the positive peak field and changed its sampled depth by less than 0.059%. Separately, halving the maximum time step from 8,640 to 4,320 s and tightening the integration tolerances produced a relative change of less than $10^{-10}$ in the temporal peak factor. These checks support the numerical stability of the reported peak-field comparisons.

Using this baseline source in the charge-continuity and Gauss equations under the conditions specified above gives positive peak fields of 6.70 MV m$^{-1}$ for proton CSDA ranges and 7.12 MV m$^{-1}$ for proton projected ranges, at depths of approximately 1.10 and 0.95 µm, respectively. Of these two results, the projected-range calculation is closer to the peak of approximately 7.1 MV m$^{-1}$ near 1 µm reported by Sana and Mishra (2026). This numerical agreement, however, does not establish which range prescription was used in the original calculation.

To examine the effect of internal redistribution during response formation, we modify the baseline source as follows:

$$S_\rho^{\text{test}}(z, t) = S_\rho^{(0)}(z, t) + f\, e\Phi_e(t)[g_{\text{departure}}(z) - g_{\text{arrival}}(z)] \tag{12}$$

where $S_\rho^{(0)}$ is the baseline source, $\Phi_e(t)$ is the energy-integrated incident electron number flux entering through the reference surface per unit area and time, with units of m$^{-2}$ s$^{-1}$, and $e$ is the positive elementary-charge magnitude. The parameter $f$ specifies the assumed mean number of internally transferred electrons per incident electron. The unit-normalized distributions $g_{\text{departure}}(z)$ and $g_{\text{arrival}}(z)$ describe departure and arrival depths; each integrates to unity over depth and has units of m$^{-1}$. Removing negative charge at departure produces a positive source correction, while adding it at arrival produces a negative correction. The two corrections cancel upon integration over depth, thus preserving the net charge supply. As an implementation check, $f = 0$ for Cases B and C recovered the baseline field exactly (maximum numerical difference of 0 V $m^{-1}$). To check charge conservation numerically, we integrated the source correction over the depth grid. At the incident-flux maximum, trapezoidal integration gave a maximum absolute residual of $1.55 \times 10^{-25}$ C $m^{-2}$ $s^{-1}$ across Cases B and C for $f = 0$–$0.25$, sampled at intervals of 0.001. The negligible residual confirms that the positive and negative source corrections cancel upon integration, leaving the depth-integrated net charge supply unchanged. This check verifies the source construction; numerical convergence of the field solution is assessed by the grid and time-step tests above. The redistribution correction is assigned to response formation, with subsequent Ohmic transport unchanged. For this example calculation, the transfer prescription is taken to be independent of incident energy.

Case A retains the baseline source without an additional source correction. Cases B and C use lognormal departure and arrival distributions defined by:

$$g(z; z_m) = [1/(z\, s_{\ln z} \sqrt{(2\pi)})] \exp[-\{\ln(z/z_m)\}^2/(2 s_{\ln z}^2)], \text{ for } z > 0 \quad (13)$$

Here, $z_m$ is the median depth, and $s_{\ln z} = 0.2$ specifies the standard deviation of $\ln(z/z_m)$. Both cases have a departure median of 10 μm, while the arrival medians are 0.3 μm for Case B and 5 μm for Case C. The transfer parameter f is varied from 0 to 0.25. These depths and transfer amounts are idealized sensitivity parameters, not inferred electron trajectories or measured reabsorption fractions. The cases distinguish redistribution across the baseline peak depth from redistribution predominantly below it. For each case, the modified source is coupled to the same charge-continuity and electrostatic equations, retaining the baseline conductivity, permittivity, initial state, and boundary conditions.

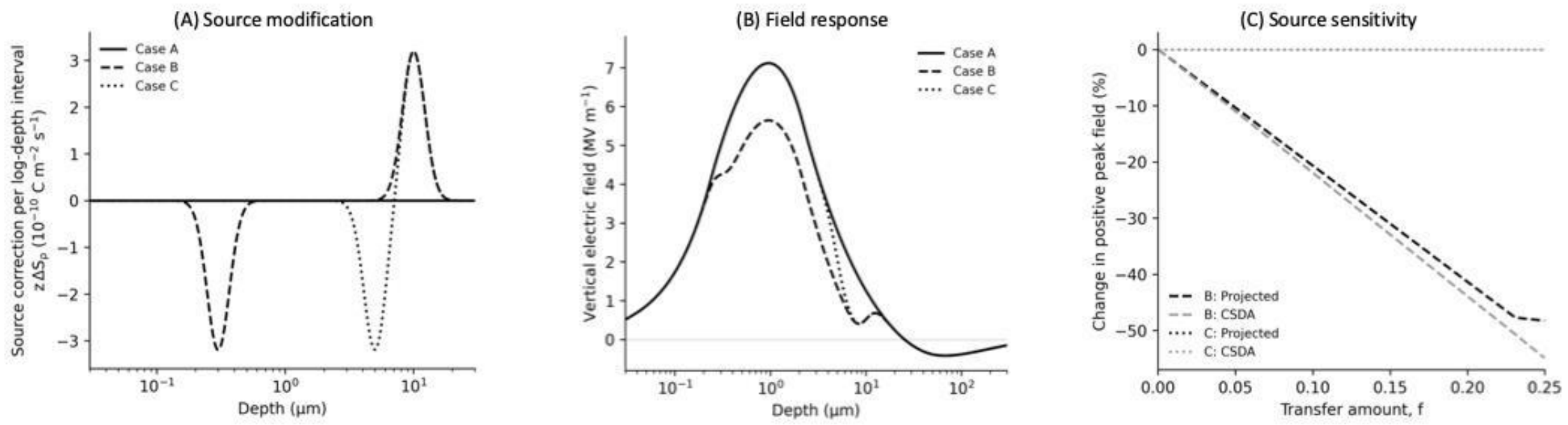

**Figure 4.** Illustrative field sensitivity to charge-conserving internal redistribution. Case A retains the baseline source; Cases B and C transfer electrons from a departure median of 10 μm to arrival medians of 0.3 and 5 μm. (A) Source corrections per unit log-depth, $z\Delta S_p$, at the incident-flux maximum (day 10; $f$ = 0.10). (B) Electric-field profiles at day 11.82 for $f$ = 0.10, using the proton projected-range baseline. (C) Positive-peak changes relative to each baseline as $f$ varies. Irradiation, electrical properties, and boundary conditions are fixed. Transfer amounts and depths are prescribed sensitivity parameters, not estimates of electron transfer in lunar regolith.

At $f$ = 0.10, Case B reduces the positive peak field from 7.12 to 5.65 MV m$^{-1}$ with proton projected ranges, and from 6.70 to 5.23 MV m$^{-1}$ with proton CSDA ranges: decreases of approximately 21% and 22%. Case C leaves the positive peak nearly unchanged while modifying the deeper field profile. The comparisons are made at the common temporal field maximum, approximately day 11.82. Figure 4 shows the source modifications, resulting field profiles, and dependence of the positive peak on $f$.

The different outcomes follow from the cumulative signed charge above the depth of interest. In Case B, redistribution adds negative charge above the baseline peak and removes it below that depth, reducing the cumulative charge that supports the positive field. In Case C, both departure and arrival occur predominantly below the peak, leaving that cumulative charge nearly unchanged. Thus, preserving the net charge supply and specifying a transfer amount do not determine the field response without information about transfer depths. These results identify a spatial constraint relevant to interpreting existing predictions, but do not demonstrate that the assumed transfers occur in lunar regolith or improve predictive accuracy. Their physical applicability requires constraints on the available carrier population and the conditions permitting transfer.

## 6 Planetary Implications

This section examines what the assessment in Section 5 implies for three successive inferences: from charge supply to macroscopic internal fields, from macroscopic fields to local dielectric breakdown, and from individual discharges to cumulative regolith weathering. For each step we state what current evidence supports and what additional evidence is required.

### *6.1 Implications for Internal Electric Fields*

Existing models suggest that energetic-particle irradiation can generate substantial subsurface electric fields where charge accumulates faster than it relaxes (Jordan et al., 2014; Sana & Mishra, 2026). Their interpretation depends partly on how particle-induced charge supply is represented. For example, Jordan et al. (2014) discuss how near-surface secondary-electron escape could reinforce positive charging, whereas the overlap between positive and negative charging layers can reduce the field between them. These considerations motivate examining whether the expected strengthening or weakening of the field persists when other charge-retention processes and material dependence are included.

Hapke's (2022) argument that secondary-electron processes and the porous structure of regolith could inhibit strong-field buildup raises a related question: do these processes reduce the depth-dependent charge separation that sustains the predicted field? As discussed in Section 5.1, evidence of electron reabsorption does not determine the direction or magnitude of that change. The calculation in Section 5.5 shows that, with net charge supply and transport fixed, the field response depends on where transferred charge departs and arrives. It does not determine whether the processes proposed by Hapke occur under lunar conditions.

At this step, current evidence supports the conclusion that predicted mean fields are conditional on unverified charge-source assumptions, but it does not determine whether existing models overestimate or underestimate those fields. A material-specific retained-charge response could be used to examine how regolith composition and spatial arrangement affect the charge profile and the depth of the predicted field peak under matched irradiation conditions. Effects on charge supply should be distinguished from the roles of conductivity in charge relaxation and permittivity in the charge–field relation, since field persistence depends on both the source and the adopted transport description.

### *6.2 Conditions for Dielectric Breakdown*

Dielectric breakdown occurs when a dielectric material loses its insulating properties under a sufficiently strong local electric field, allowing rapid current flow. Existing studies have proposed that large SEP events may produce conditions capable of inducing breakdown within lunar regolith (Jordan et al., 2014, 2015; Sana & Mishra, 2026). The failure conditions depend on the material and local electrical environment, rather than on incident particle fluence alone (Jordan et al., 2015).

Hapke (2022) questions whether the charge accumulation required for these predictions can be sustained in porous regolith, whereas MacLeod et al. (2023) observed electrical discharges and associated damage in single-crystal silicate sections irradiated with 30 keV electrons at room temperature in high vacuum. These findings address different steps in the connection between charging and failure. The experiments support the possibility of electrical damage under the tested conditions, but do not show whether a porous regolith volume retains the charge distribution needed to reach comparable failure conditions. In particular, charge exchange among grains and pore walls must be considered when applying sample-scale results to the bulk regolith. The experimental evidence therefore does not resolve the question of lunar charge accumulation raised by Hapke.

Section 5.5 demonstrates that uncertainty in the spatial distribution of charge supply can change the predicted positive peak field even when the net charge supply is preserved. The implication for a threshold-based breakdown prediction depends on whether this change moves the field across the adopted threshold. If it does, the prediction is sensitive to the charge-source description; if not, the source variations examined do not change that prediction within the adopted criterion.

A depth-dependent macroscopic field does not resolve the local fields responsible for failure. Irregular grain geometries and interfaces between materials with contrasting dielectric properties have been discussed as potential sites of field concentration (Andres, Timoshkin,

Jirestig, & Stallknecht, 2001; Andres, Timoshkin, & Soloviev, 2001; Bahder et al., 1982; Hara & Okubo, 1998; Jordan et al., 2015; Lisitsyn et al., 1998). These studies, most of which concern engineered insulators and ores rather than regolith, provide mechanisms to consider but do not quantify their importance under lunar regolith conditions. Identifying where breakdown occurs requires local fields around grains and pores and failure properties appropriate to the materials, temperatures, and field durations considered. Changes in the predicted macroscopic field therefore cannot automatically be translated into corresponding changes in breakdown probability. At this step, current evidence supports breakdown as a possible mechanism under laboratory conditions; it does not support quantitative breakdown rates or locations in lunar regolith.

### *6.3 Possible Contribution to Regolith Weathering*

Electrical damage in irradiated silicate minerals demonstrates a material-modification pathway under the tested conditions (MacLeod et al., 2023). It does not independently validate a lunar retained-charge profile or determine discharge rates in permanently shadowed regions. A weathering assessment additionally requires constraints on local failure, damage per discharge, and cumulative exposure.

Repeated breakdown has been proposed as a mechanism of regolith weathering, potentially modifying grain size, structure, and optical properties (Jordan et al., 2015). Assessing this contribution requires relating the energy released and material affected per discharge to the changes accumulated under repeated exposure. The time required to reach a prescribed field threshold should therefore be distinguished from a recurrence time, which also depends on how a discharge changes the charge distribution and material. Changes in grain size, porosity, or mineral interfaces could in turn alter subsequent charging and failure, so the effects of repeated discharges cannot be assumed to remain constant.

Applying laboratory damage observations to lunar regolith requires accounting for differences in composition, grain–pore structure, and temperature. The extent and depth of material modification must also be related to the material sampled by lunar observations, while the experimental irradiation history must be distinguished from the exposure history over which lunar changes accumulate. Similarities between experimental damage and lunar features would support further comparison, but would not uniquely identify breakdown as their cause. A contribution of dielectric breakdown to lunar space weathering therefore remains a hypothesis that current evidence permits but does not yet test quantitatively; improved retained-charge constraints would strengthen its charging basis, while its cumulative contribution would depend on the additional links between discharge, damage, and lunar observations.

## 7 Conclusions

This review assessed what existing models, emission and dust-charging studies, particle-resolved calculations, and silicate irradiation experiments constrain about charge retention in lunar regolith. Four conclusions follow.

*First, measured and calculated quantities carry different information.* Particle stopping and energy deposition do not uniquely determine the signed charge remaining within regolith. Emission measurements constrain electron loss across the measured boundary, while evidence of reabsorption indicates that charge can be redistributed internally without leaving the selected volume. A low emitted yield therefore does not justify neglecting internal redistribution, nor does evidence of reabsorption demonstrate that depth-dependent charge separation is reduced. In the literature reviewed, no measurement constrains the depth-dependent retained charge in lunar soils or simulants under SEP- or GCR-relevant irradiation.

*Second, the principal studies in the lunar breakdown debate address different steps and must be evaluated together rather than as competing answers to a single question.* Jordan et al. (2014) and Sana and Mishra (2026) predict strong subsurface fields under prescribed charge-source and transport assumptions. Hapke (2022) questions whether the required charge accumulation can be sustained in porous regolith, while MacLeod et al. (2023) demonstrate electrical damage in silicate crystals under laboratory conditions. None of these studies alone determines whether lunar regolith develops the charge distributions required for breakdown.

*Third, the illustrative calculation shows that the same net charge supply can produce different field profiles when internal transfer occurs between different depths.* Within the tested assumptions, transfer across the baseline peak depth reduces the positive peak, whereas transfer

predominantly below it leaves that peak nearly unchanged while altering the deeper field. Constraints on internal transfer must therefore include departure and arrival depths, not only the amount transferred. The calculation does not show that such transfer occurs in lunar regolith.

*Fourth, mean-field prediction, local breakdown, and cumulative weathering each require distinct evidence.* The average signed excess charge per unit depth per incident particle, defined for specified species, energy, material, irradiation, electrical-state, and evaluation conditions, provides a conditional input for mean-field calculations, but does not resolve local fields around grains and pores. Predicting breakdown additionally requires local fields and failure properties appropriate to cryogenic, porous regolith, and assessing weathering further requires discharge recurrence, damage per discharge, and exposure history.

The highest priority for future work is therefore not further refinement of mean-field models alone, but measurements that jointly constrain the boundary charge budget, the spatial separation of retained charge, and its temporal relaxation in lunar-relevant materials under cryogenic, high-vacuum conditions. Combined with particle-resolved modelling and independent tests of the resulting field predictions, such measurements would allow the charge-source assumptions of existing models to be evaluated directly.

## Open Research

The NIST ESTAR and PSTAR range data used in Section 5.5 are available from the NIST Physical Measurement Laboratory (Berger et al., 2005). The sensitivity calculation was implemented and checked using Python 3.12.14 with NumPy 2.3.5 and SciPy 1.17.0. Numerical plots were generated using Matplotlib, and schematic illustrations and final figure layouts were prepared using Microsoft PowerPoint.

## Conflict of Interest Disclosure

The authors declare there are no conflicts of interest for this manuscript.